\documentstyle[preprint,prb,aps]{revtex}
\tightenlines
\begin{document}
\draft
\title{Far-infrared excitations in a quantum antidot at finite magnetic
fields}
\author{Agust\'{\i} Emperador, Mart\'{\i} Pi, and Manuel Barranco}
\address{Departament d'Estructura i Constituents de la Mat\`eria,
Facultat de F\'{\i}sica, \\
Universitat de Barcelona, E-08028 Barcelona, Spain}
\author{Enrico Lipparini}

\address{Dipartimento di Fisica, Universit\`a di Trento, and INFM,
sezione di Trento.  38050 Povo, Italy}
\author{Lloren\c{c} Serra}
\address{Departament de F\'{\i}sica, Facultat de Ci\`encies,\\
Universitat de les Illes Balears, E-07071 Palma de Mallorca, Spain}
\date{\today}

\maketitle

\begin{abstract}

We have investigated the far-infrared dipole modes of a
quantum antidot submitted to a perpendicularly applied 
magnetic field $B$.
The ground state of the antidot is described within
local spin-density functional theory, and the spectrum within
time-dependent local spin-density functional theory.
The results are compared with those corresponding
to a quantum dot of similar electronic surface density. 
The method is able to reproduce two of the more salient
experimental features, namely that main bulk and edge modes have the
same circular polarization, and that the negative  $B$
dispersion edge branch oscillates, having minima at the
$B$ values corresponding to fully occupied Landau levels. It fails,
however, to yield the unique feature of short-period antidot
lattices that the energy of the
edge magnetoplasmon approaches the cyclotron frequency for small
$B$. The existence of anticyclotron polarized bulk modes is 
discussed, and a detailed account of the dipole spin mode 
is presented.

\end{abstract}


%

\section{Introduction}

The far-infrared (FIR) spectroscopy of laterally confined superlattices
in the two dimensional electron gas (2DEG) made of  holes
surrounded by electrons, called antidots,
 has uncovered\cite{Ker91,Zha92,Lor92}
a very peculiar behavior of the collective spectrum of these
systems. It has been found\cite{Ker91,Zha92} that
it consists of a high frequency branch which starts either with a
negative magnetic field $(B)$ dispersion in the case of high electronic
surface density $n_s$, or with a rather flat $B$ dispersion at 
low $n_s$, that eventually converges to the cyclotron energy at 
high $B$'s,
plus a low frequency branch which at high $B$'s corresponds to
the usual edge magnetoplasmon, but that approaches the cyclotron
frequency for small magnetic fields. Anticrossing of the modes
appears as $B$ increases. 
It is worth to recall that the mentioned behavior of the high frequency
branch is also a feature of the FIR spectrum of quantum
rings\cite{Dah93,Emp99}. 

A study employing circularly
polarized radiation\cite{Bol95} has shown that both edge and bulk
magnetoplasmons in antidots exhibit the same circular polarization,
in contradistinction with what is found in quantum
dots\cite{Sik89,Dem90}. Interestingly, a very recent 
experiment\cite{Hoc99} has detected  weak bulk modes 
anticyclotron-like polarized, whose existence had been predicted some
time ago\cite{Wu93,Mik95}.

Another difference with dots appeared after a
careful analysis of the edge magnetoplasmon\cite{Bol96}:
the frequency of this mode shows a conspicuous oscillation with $B$
that has  maxima at fully occupied Landau levels (even filling
factors $\nu$) in the case of dots, and at half-filled Landau levels
(odd $\nu$'s) in the case of antidots.

Several theoretical descriptions of  FIR modes in antidots
have been given in the past. Some are  based on
hydrodynamical\cite{Wu93,Fes93} or  classical electrodynamics
\cite{Mik95,Mik96} models, which either do not take into
account the periodicity of the antidot array\cite{Fes93}, or which
take it into account either in a circularly symmetric Wigner-Seitz
approximation\cite{Wu93}, or incorporate the parameters of
the experimental short-period antidot lattices without further
approximations\cite{Mik96}.

Despite that all these models have somehow succesfully reproduced the
FIR absorption in antidots, the interest in achieving a more microscopic
description of the ground state (gs) and FIR response still
remains. Quantum mechanical (Hartree) models have been set up which take
into account the periodicity of the lattice using confining
potentials of different complexity\cite{Hua93,Gud96,Gud98}.
The difficulties
inherent to the handling of many single particle (sp) wave functions
and realistic confining potentials which reflect the actual
antidot lattice have hampered these microscopic methods to achieve a
quantitative description of the FIR absorption in antidots, although
some features of the process are qualitatively 
described\cite{Hua93,Gud98}.

Recently, we have applied a local spin-density functional (LSDFT)
method to calculate the structure  of isolate
antidots at $B=0$, together with a  sum rule approach to describe
their FIR spectrum\cite{Emp98}. 
Our aim here is to extend these calculations 
to the case of finite magnetic fields and to present an account of
the gs and dipole response of an antidot within the frame of
LSDFT and its time-dependent generalization (TDLSDFT). 
We recall that these formalisms
incorporate the spin degree of freedom and allow to
selfconsistently include  exchange and correlation effects
in the study of the gs  and the response of the system, whose
importance in the description of 2D electronic structures is
nowadays well established.

Rather than isolated antidots, the systems described in this work 
should be considered as representing long-period antidot lattices.
Although no much differences are expected to appear
between  the ground state structure of an isolated antidot and
that of an antidot in a long-period array, apart from the obvious 
changes at the border of the unit cell, bulk
magnetoplasmons are qualitatively different in both 
cases\cite{Mik95}.

\section{The LSDFT and TDLSDFT approaches}

We have modeled an antidot of radius $R$ in a 2DEG of surface
density $n_s$ by a positive jellium background of density
$n_J(r) = n_s\,\Theta (r - R)$. 
The gs of the antidot is
obtained solving the Kohn-Sham (KS) equations. 
The problem is simplified by the
imposed circular symmetry, which allows one to write the 
single particle (sp) wave functions as $\phi_{nl\sigma}(r,\theta)=
u_{nl\sigma}(r) e^{-\imath l \theta}$ with 
$l =0, \pm 1, \pm 2, ...$, being $-l$ the sp orbital angular momentum.

We have used effective atomic units defined by
$\hbar=e^2/\epsilon=m=$1, where 
$\epsilon$ is the dielectric constant,
and $m$  the electron effective mass. In units of the bare
electron  mass $m_e$ one has $ m = m^* m_e$.
In this system of units, the length unit is the effective
Bohr radius $a_0^* = a_0\epsilon/m^*$,
and the energy unit is the effective Hartree $H^* = H  m^*/\epsilon^2$.
In the numerical applications we have always considered
GaAs, for which  we have taken $\epsilon$ = 12.4,  $m^*$ = 0.067, and
$g^*=-$0.44. This yields $a^*_0 \sim$ 97.94 ${\rm \AA}$ 
and $H^*\sim$ 11.86 meV $\sim$ 95.6 cm$^{-1}$.
The  Bohr magneton  is  $\mu _B=\hbar e/2 m_e c$, and
the cyclotron frequency is $\omega_c = e B/mc$.

The radial KS equations read

\begin{eqnarray}
& & \left[-\frac{1}{2} \left( \frac{d^2}{d r^2}
+ \frac{1}{r} \frac{d}{d r} - \frac{l^2}{r^2} \right)
-\frac{\omega_c}{2} l+ \frac{1}{8} \omega_c^2 r^2
- V^+(r) \right.
\nonumber
\\
& &
\label{eq1}
\\
&+& \left. V^H + V^{xc} +(W^{xc}+{1\over 2}g^*\mu_B B)\,
\eta_\sigma \right]
u_{n l \sigma} =
\epsilon_{n l \sigma} u_{n l \sigma} \,\, ,
\nonumber
\end{eqnarray}
where $\eta_\sigma=+1 (-1)$ for $\sigma=\uparrow (\downarrow)$, 
$V^H=\int{d\vec{r}\,'n(\vec{r}\,')/|\vec{r}-\vec{r}\,'|} $
is the Hartree potential, and $V^{xc}={\partial
{\cal E}_{xc}(n,m)/\partial n}\vert_{gs}$ and
$W^{xc}={\partial
{\cal E}_{xc}(n,m)/\partial m}\vert_{gs}$
are the  variations of the exchange-correlation
energy density ${\cal E}_{xc}(n,m)$ written in terms of the electron density 
$n(r)$ and of the local spin magnetization 
$m(r)\equiv n^{\uparrow}(r)-n^{\downarrow}(r)$ taken at the ground state.
${\cal E}_{xc}(n,m)$ has been built from the 2DEG calculations
of Tanatar and Ceperley\cite{Tan89} at zero and full spin polarization,
using the Von Barth and Hedin\cite{vBa72} interpolation formula 
adapted to 2D.
The jellium potential $V^+(r)$ is analytical\cite{Emp98}:

\begin{equation}
V^+(r)  =  4 n_s \times \left\{ \begin{array}{ll}
R_{\infty}\,{\bf E}(r/R_{\infty})
-R \,{\bf E}(r/R) &  r < R  \\
R_{\infty}\,{\bf E}(r/R_{\infty}) - r\, {\bf E}(R/r)
+ r \,[ 1 - (R/r)^2\,] \,{\bf K}(R/r)
 &     r > R  \,\, ,
                \end{array}
\right.
\label{eq2}
\end{equation}
where {\bf K} and {\bf E} are the complete
elliptic integrals of first and second kind, respectively
\cite{Gra80}, and $R_{\infty}$ represents a large $r$ value.
In practice, it
is the largest-$r$ used in the gs calculation.
The KS differential equations 
have been solved without expanding the sp wave
functions in a necessarily truncated basis of  Landau
orbitals. Our iterative method  works for weak and strong $B$ fields
as well, for which the effective potential is
very different. It has the advantage of avoiding to study how
the results depend on the size of the basis. As  in Ref. 
\onlinecite{Pi98}, we have worked at a small but finite
temperature (0.1 K).

Physically acceptable solutions have to be regular at $r = 0$, 
and have to be bound due to the $\omega_c^2$ term in the KS equation. 
Far from $R$ they behave  as\cite{Foc28,Dar30}

\begin{equation}
u_{nl\sigma}(r) \sim
 \left(\frac{r}{{\cal L}}\right)^{|l|}
e^{-(r/2{\cal L})^2} L^{|l|}_n\left(\frac{r^2}{2{\cal L}^2}
\right) \,\,\, ,
\label{eq3} 
\end{equation}
where ${\cal L} \equiv \sqrt{\hbar/m\omega_c}$ is the magnetic length
and $L^{|l|}_n$ is a generalized Laguerre polynomial\cite{Abr70}.

Once the KS ground state has been obtained, we determine the
dipole longitudinal response employing linear-response 
theory\cite{Ser99}. We sketch here how the method is applied.

For independent electrons in the KS mean field,
the variation $\delta n^{(0)}_{\sigma}$ induced in the spin
density $n_{\sigma}$  ($\sigma\equiv\uparrow,\downarrow$) by an external
field $F$, whose non-temporal dependence we denote as
$F=\sum_{\sigma}f_{\sigma}(\vec{r})\,|\sigma\rangle\langle\sigma|$,
can be written as \cite{Wil83}
\begin{equation}
\delta n^{(0)}_{\sigma}(\vec{r},\omega) =
\sum_{\sigma'}\int d\vec{r}\,'\chi^{(0)}_{\sigma\sigma'}
(\vec{r},\vec{r}\,';\omega)f_{\sigma'}(\vec{r}\,')\; ,
\label{eq4}
\end{equation}
where
$\chi^{(0)}_{\sigma\sigma'}$ is the KS spin density correlation function.
In this limit, the frequency $\omega$ corresponds to the
harmonic time-dependence of the external field $F$ and  of
the induced $\delta n^{(0)}_\sigma$. Eq.\ (\ref{eq4}) is a 2$\times$2
matrix equation in the two-component Pauli space.
In longitudinal response theory, $F$ is diagonal in this space,
and its diagonal components are written as a vector
$F\equiv\left(
\begin{array}{c} f_\uparrow\\ f_\downarrow\end{array}
\right) $.
We consider  external dipole ($L$ = 1) fields of the kind
\begin{equation}
F^{(n)}_{\pm 1}=  f(r) \,e^{ \pm \imath \theta}
\left(\begin{array}{c} 1\\ 1\end{array}\right)
\,\,\, {\rm and} \,\,\,
F^{(m)}_{\pm 1}   =    f(r)\, e^{\pm \imath  \theta}
\left(\begin{array}{c} 1\\ -1\end{array}\right)
\label{eq5}
\end{equation}
which cause, respectively, charge and spin density modes
of dipole type. To distinguish the induced densities in each
excitation channel, we label them with an additional superscript as
$\delta n^{(0,n)}_\sigma$ or  $\delta n^{(0,m)}_\sigma$.

The TDLSDFT induced densities are obtained solving the integral
equations
\begin{eqnarray}
\delta n^{(A)}_{\sigma}(\vec{r},\omega) =
\delta n^{(0,A)}_{\sigma}(\vec{r},\omega)+
\sum_{\sigma_1\sigma_2}\int d\vec{r}_1d\vec{r}_2\,
\chi^{(0)}_{\sigma\sigma_1}(\vec{r},\vec{r}_1;\omega)
K_{\sigma_1\sigma_2}(\vec{r}_1,\vec{r}_2)\,
\delta n^{(A)}_{\sigma_2}(\vec{r}_2,\omega) \,\,\, ,
\label{eq6}
\end{eqnarray}
where either $A=n$ or $A=m$, and
the kernel $K_{\sigma\sigma'}(\vec{r},\vec{r}\,')$
is the electron-hole interaction.

Equations (\ref{eq6}) have been solved as a generalized matrix
equation in coordinate space. Considering
angular decompositions of  $\chi_{\sigma\sigma'}$ and
$K_{\sigma\sigma'}$ of the kind
$K_{\sigma\sigma'}(\vec{r},\vec{r}\,')=
\sum_l K^{(l)}_{\sigma\sigma'}(r,r') e^{i l(\theta -\theta')}
$, it is enough to solve them for $l = 1$
because only this term couples to the external dipole field. 

When a magnetic field is perpendicularly applied to the antidot,
the $\pm 1$ modes are not degenerate and two excitation branches 
may appear, each having {\em in principle} a different
circular polarization, i.e., carrying an orbital angular momentum
$\Delta L_z=\pm 1$, where $L_z$ is that of the gs.
In contradistinction with the quantum dot (QD) case, we will see 
that this is not so for antidots, for which the more
intense peaks in both branches have the same polarization.

The induced  charge or magnetization densities corresponding to
density and spin responses are given by
$\delta n^{(A)}=\delta n^{(A)}_\uparrow+
\delta n^{(A)}_\downarrow$ and
$\delta{m}^{(A)}=\delta n^{(A)}_\uparrow-
\delta n^{(A)}_\downarrow$.
>From them, the dynamical polarizabilities in the density and spin
channels are respectively given by
\begin{eqnarray}
\alpha_{nn}(\omega)&=&
\int{dr \,r\, f(r)\,  \delta n^{(n)}(r)}\nonumber\\
\alpha_{mm}(\omega)&=&
\int{dr\, r\, f(r)\, \delta m^{(m)}(r)} \; .
\label{eq7}
\end{eqnarray}
In this expression $\delta n^{(n)}$ has to be
understood as the
charge density induced by a spin-independent operator, and
$\delta m^{(m)}$ as the spin density magnetization induced by
a spin-dependent operator. Within longitudinal response theory,
cross-channel induced densities such as $\delta m^{(n)}$ or
$\delta n^{(m)}$ may also appear\cite{Ser99}. The above
expressions hold for each $\pm 1$ circular polarization of the $F$
operators defined in Eq. (\ref{eq5}). 
Taking into account both possibilities, we
define $\alpha_{AA}^{(1)}(\omega) \equiv \alpha_{AA}(1,\omega)
+ \alpha_{AA}(-1,\omega)$, whose imaginary part is proportional
to the strength function  $S^{(1)}_{AA}(\omega)=
{\rm Im}[\alpha^{(1)}_{AA}(\omega)]/\pi$. The peaks appearing in the
strength functions are  dipole charge density  (CDE) or
spin density excitations (SDE) caused
by the external field. Analogously, the peaks appearing
in the strength function which results from using in Eq. (\ref{eq7})
the KS  induced densities $\delta n^{(0,A)}_{\sigma}$ instead of
the correlated ones $\delta n^{(A)}_{\sigma}$, 
correspond to dipole single particle excitations (SPE).

We have used for the $f(r)$ function entering Eq. (\ref{eq5}) two
different choices. One is the standard dipole operator $f(r)= r$, and
the other one is $f(r)= 1/r$. As indicated in Ref. \onlinecite{Emp98}, 
the latter choice is inspired in the small-argument expansion of
the irregular Bessel function $Y_1(qr)$. The relevance of this operator 
in the present context is that it mostly causes edge excitations.
This will be futher discussed in the next Section. The situation is
analogous to that faced in the description of surface modes of 3D 
cavities in metals, see  Ref. \onlinecite{Ser92} for a thorough 
discussion.

\section{Results and discussion}

As a case of study we have considered an antidot of $n_s = 0.25$ 
$(a^*_0)^{-2}$ and $R = 7.5$ $a^*_0$. It roughly corresponds
to one of the systems studied by Zhao et al \cite{Zha92}
having $n_s = 2.6 \times 10^{11}$ cm$^{-2}$ and the same area. We
recall that the antidot arrays studied in this reference have
a very short period, whereas we have taken for $R_{\infty}$ 
defined in Eq. (\ref{eq2}) a mean value of 30 $a^*_0$, larger for
large $\nu$'s, and smaller for small $\nu$'s.
We have employed $B$ values yielding integer filling factors 
$\nu = 2 \pi\, n_s \,{\cal L}^2$
in the range $1 \leq \nu \leq 10$ which corresponds to
$10.8 \geq B \geq 1.08$ T.

\begin{figure}
\includegraphics{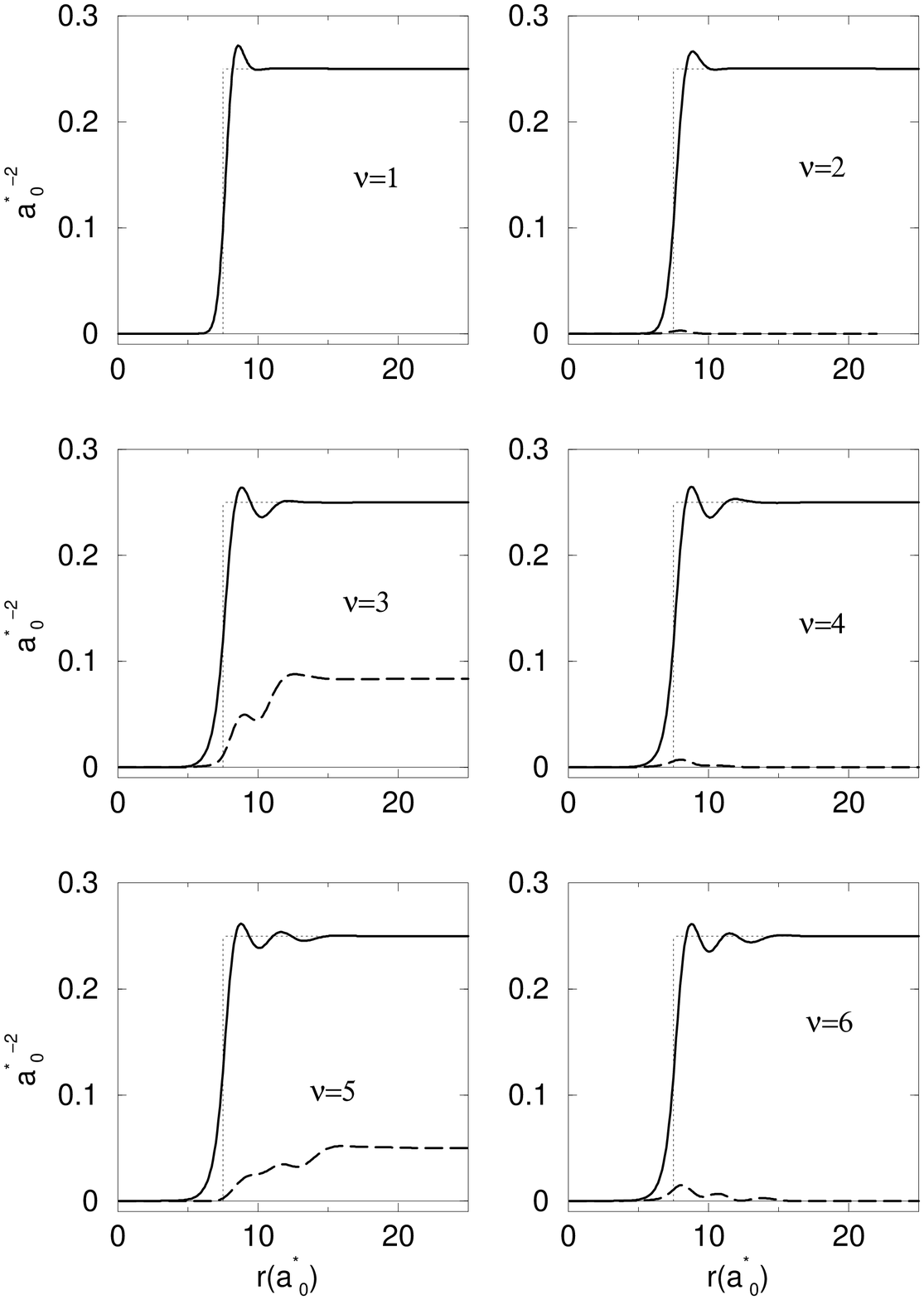}
\vspace*{19.5cm}
\caption{ 
Electron (solid lines) and spin magnetization 
(dashed lines) densities as a function of
$r$  for  an antidot of $R$ = 7.5 $a_0^*$ and $n_s$ = 0.25
(a$^*_0$)$^{-2}$ and several filling factors.
Also shown is the jellium density (dotted lines).
}
\label{fig1}
\end{figure}

Figure \ref{fig1} shows the particle $n(r)$ and spin magnetization 
densities $m(r)$ as a function of $r$ for $\nu = 1$ to 6. 
Away from $R$, the electron density
approaches $n_s$, and either $m(r)$ is zero if $\nu$ is even, or
the local spin polarization
$\xi(r) \equiv m(r)/n(r)$ equals $1/\nu$. The number
of sp orbitals needed to numerically achieve this limit increases
with $\nu$, making it rather involved  to obtain  high-$\nu$ 
ground states. For instance, some 2500 occupied sp levels have been
used to model the $\nu= 10$ gs.

\begin{figure}
\includegraphics{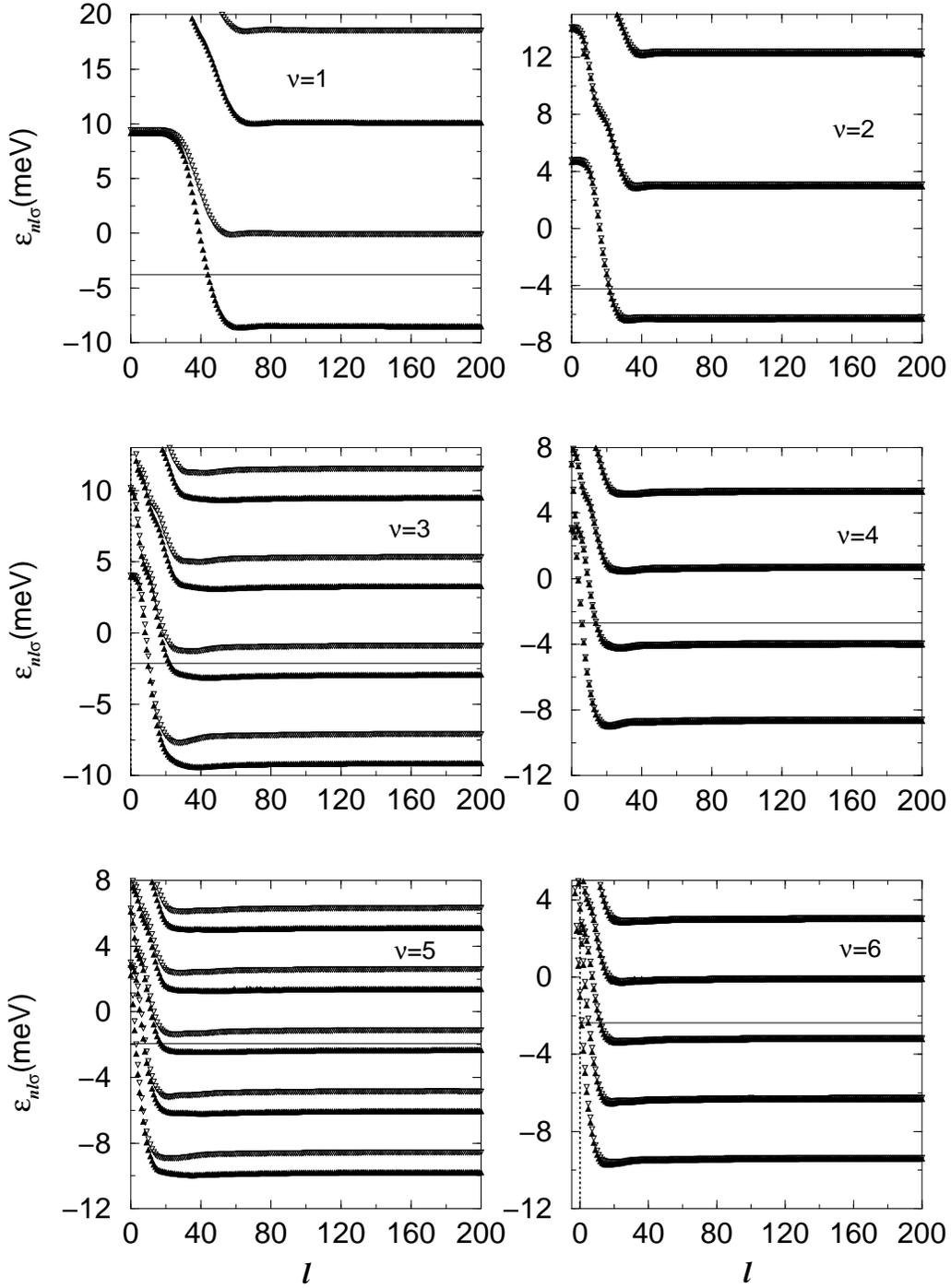}
\vspace*{19.5cm}
\caption[]{
Single particle energies $\epsilon_{nl\sigma}$ as a function of
$l$ for the configurations of Fig. \ref{fig1}.
The horizontal line represents the electron chemical potential.
Full upright triangles represent $(M,\uparrow)$ bands, and empty
downright triangles represent $(M,\downarrow)$ bands.
}
\label{fig2}
\end{figure}

Figure \ref{fig2} shows the sp energies $\epsilon_{nl\sigma}$ as
a function of $l$ for the same configurations as in Fig. 
\ref{fig1}. In the bulk, $\epsilon_{nl\sigma}$ are arranged into
Landau bands characterized by the index $M \equiv n +(|l|-l)/2$ and
the value of $\sigma$. The filling factor represents the
number of occupied Landau bands, each of them labeled as
$(M,\uparrow {\rm or} \downarrow)$. It is worth noticing that since
we have taken $B$ in the positive $z$ direction and the sp orbital
angular momentum is written as $-l$, most occupied sp levels have
$l \geq 0$, and that for a given $M$ the $(M,\uparrow)$ band lies below
the $(M,\downarrow)$ one. 

These bands are almost  degenerate when $\nu$ is even
 because of the smallness of the Zeeman energy
and because $m(r) \simeq 0$ in the gs. When $\nu$ is odd, the 
spin dependence of the exchange-correlation energy ${\cal E}_{xc}$
is responsible of the large gap between the
$(M,\uparrow \downarrow)$ bands\cite{Lip99}. Defining the $M$-th Landau
level as the set of $(M,\uparrow)$ and $(M,\downarrow)$ bands
with $M=0$, 1, 2, ..., we  see from Fig. \ref{fig2} that
even $\nu$ values correspond to completely filled Landau levels
configurations, and odd $\nu$ values to configurations in which one
Landau level is half-filled. If $M$ is not too small,
the $M$-th level is made of sp states having $n= M$. As in QD's,
this rule is violated by a few sp states per Landau level at most,
see for instance Fig. 1 of Ref. \onlinecite{Lip99}, and the $\nu = 6$
panel of Fig. \ref{fig2}.

\begin{figure}
\includegraphics{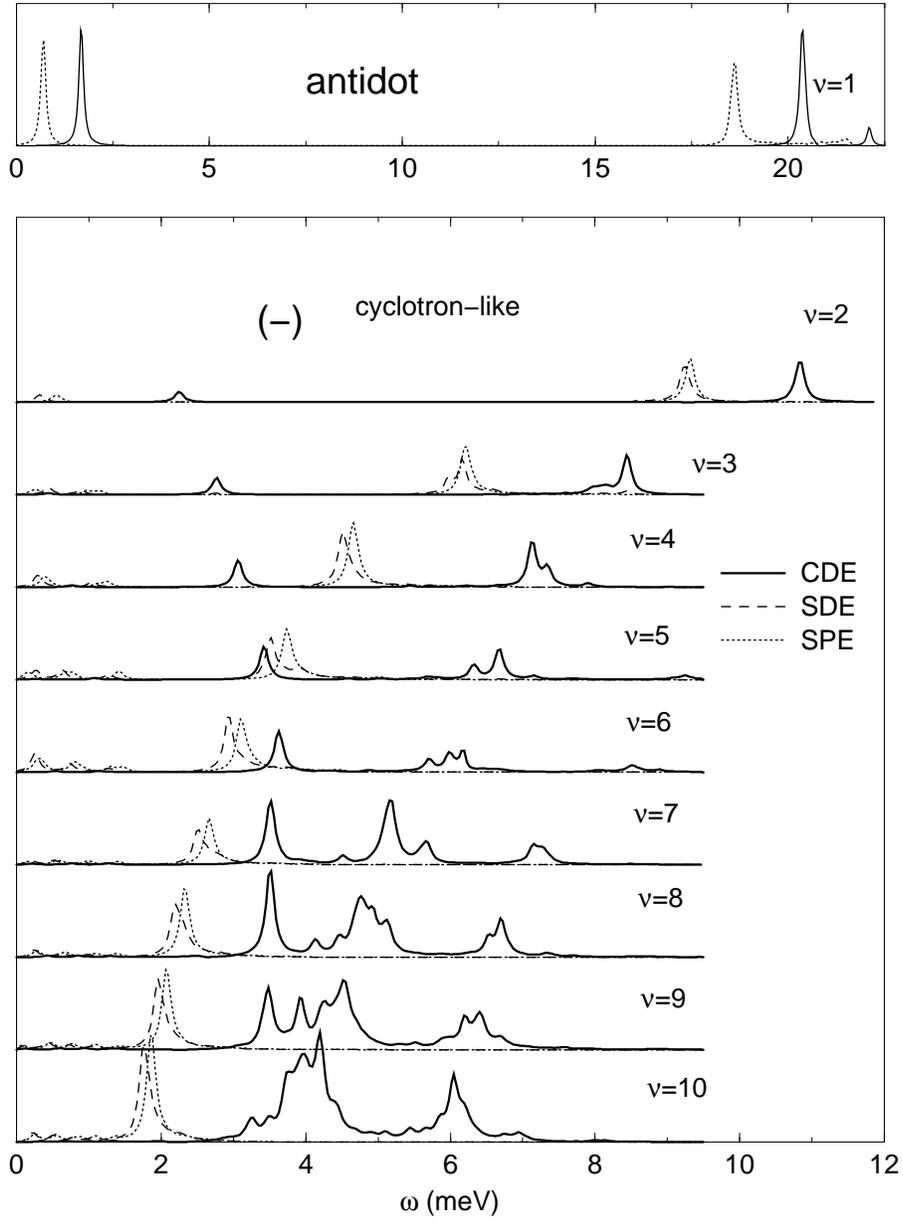}
\vspace*{19.5cm}
\caption[]{
Charge, spin, and single particle antidot dipole strength functions
in arbitrary units.
All displayed modes are cyclotron-like polarized ($-$). Notice the
different energy scale of the $\nu=1$ responses.
}
\label{fig3}
\end{figure}

The dipole FIR response has been obtained as outlined in Sect. II.
Besides evaluating the occupied Landau levels, two more empty levels 
have been calculated, which allows an accurate description 
of modes  up to $2\,\omega_c$. 
Figure \ref{fig3} shows the charge, spin,
and single electron strength functions corresponding to
the dipole operator.
All peaks displayed in this figure are of cyclotron-like type,
i.e., they are excited by the $e^{+\imath\theta}$
component of the $F$ operators in Eq. \ref{eq5}. This makes
possible the transfering of strength among them, an effect
that has been experimentally observed\cite{Ker91,Zha92}.
We remark in passing that
the fact that all these peaks originate from the same
excitation operator makes the sum rule approach 
\cite{Emp98,Lip97} unsuitable to describe their 
excitation energies. 

Our sign convention implies that at $B\neq0$
the gs has a {\em negative} total orbital angular momentum $L_z$.
Thus, these excitations
{\em decrease} in one unit the absolute value of $L_z$; whenever
needed, we shall label cyclotron-like modes with a $(-)$ sign.
Bulk magnetoplasmons in QD's also bear this
character, whereas edge magnetoplasmons do not
(see for instance Refs. \onlinecite{Ser99,Gud91}).
Indeed, edge magnetoplasmons in QD's are excited by the 
$e^{-\imath\theta}$ component of
the dipole field and correspond to excitations which
{\em increase} in one unit the absolute value of $L_z$. Whenever
needed, we shall label the anticyclotron-like modes
with a $(+)$ sign. The cyclotron-like character of edge
magnetoplasmons in antidots has its microscopic origin in the upwards
bending of the Landau levels for {\em small} $l$ values, whereas the
anticyclotron-like character of the edge magnetoplasmons in dots comes
from the upwards bending of the Landau levels for {\em large} $l$
values.

As in quantum dots, SDE's are weakly redshifted with respect to SPE's
because of the attractive character of the rather small
exchange-correlation vertex corrections,
whereas CDE's are strongly blueshifted because of the repulsive 
character of the intense direct Coulomb interaction. 
In contradistinction with the
QD case\cite{Ser99}, no clear signature of a collective longitudinal
spin edge magnetoplasmon appears at small $B$ values:
the edge spin mode is rather fragmented and weak. We recall that
at full magnetizacion {\em longitudinal} spin and density 
responses coincide\cite{Ser99} and only spin modes of 
transverse type, not studied here, can be excited\cite{Lip99}.

\begin{figure}
\includegraphics{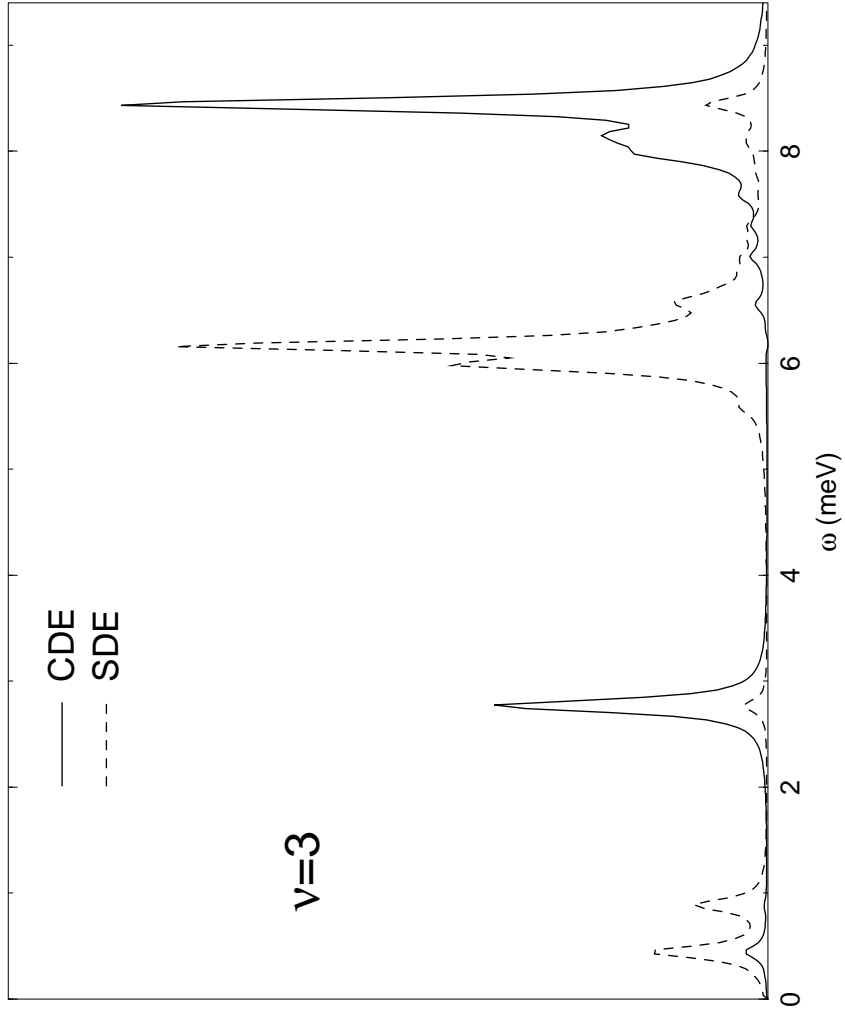}
\vspace*{19.5cm}
\caption[]{
Charge and spin antidot dipole strength functions at $\nu=3$ in 
arbitrary units illustrating the coupling between the strengths 
at high spin magnetizations.
}
\label{fig4}
\end{figure}

Another interesting feature of longitudinal response theory is that
spin and charge density modes are generally coupled if the gs has a
large spin magnetization, as it happens for instance at $\nu=3$. This
means that spin dependent probes may excite charge density 
modes. Conversely, spin independent probes may excite spin density
modes. This coupling has been experimentally detected in
QD's using Raman spectroscopy\cite{Sch98}, and has been theoretically
addressed\cite{Ser99} within TDLSDFT.
 For an antidot the effect is  more
marked, since $\xi$ reaches $1/\nu$ instead of a smaller effective
value as it happens in QD. This can be seen in Fig. \ref{fig4}.

\begin{figure}
\includegraphics{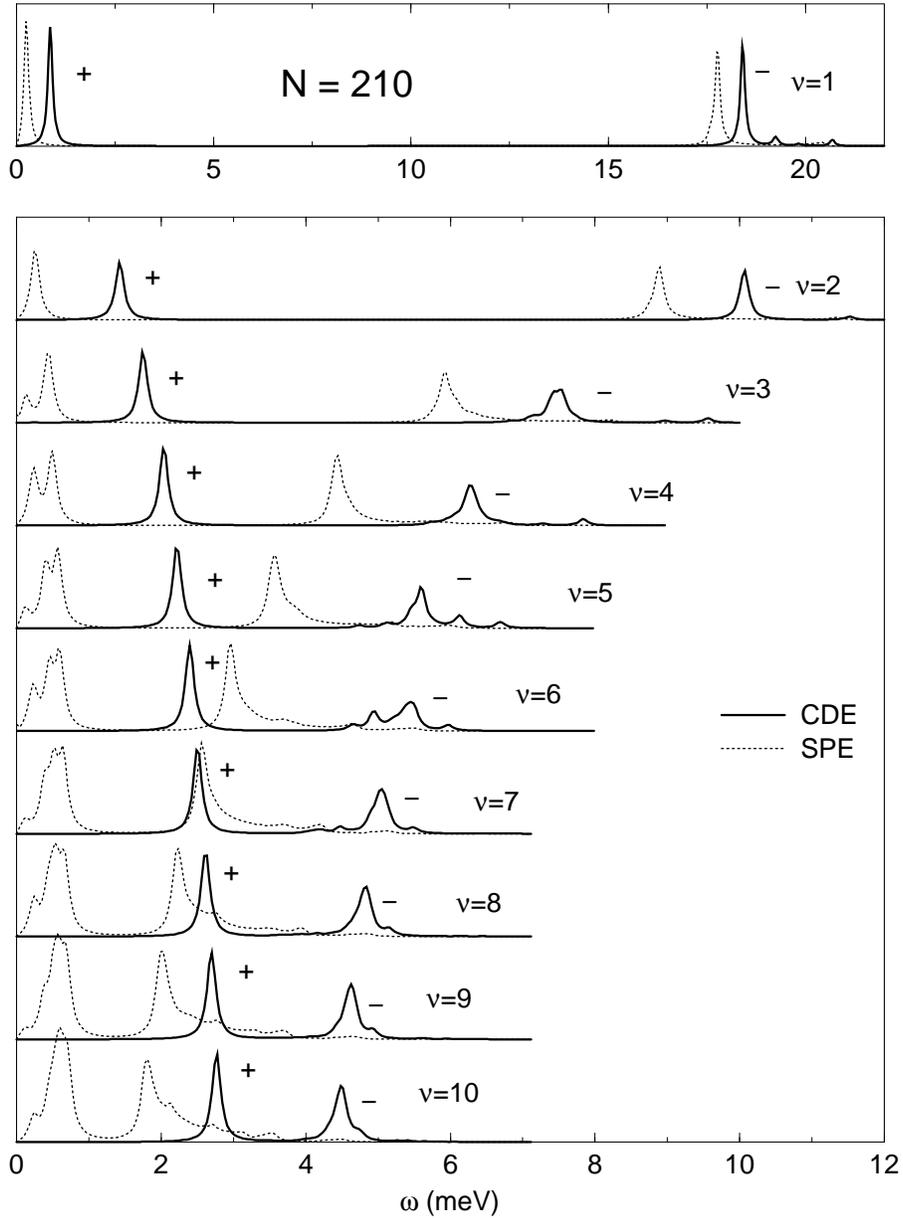}
\vspace*{19.5cm}
\caption[]{
Charge and single particle dipole stregth functions in 
arbitrary units corresponding to the $N= 210$ dot 
described in the text. Cyclotron-like polarized ($-$) bulk
modes and anticyclotron polarized ($+$) edge modes 
appear in the responses. Notice the  different energy scale
of the $\nu=1$ responses.
}
\label{fig5}
\end{figure}

It is worthwhile to compare  antidot with dot results.
For this sake, we show in Fig. \ref{fig5} the dipole
charge density response of a QD made of $N=210$ electrons confined by
the potential created by  a jellium disk of radius $R\sim 16.7 \,a^*_0$.
Outside the edge region, this system  has a  surface density of
$\sim 0.24 (a^*_0)^{-2}$, similar to our case-of study antidot,
and it has been thoroughly studied in Ref. \onlinecite{Pi98}.
For this dot the $1 \leq \nu \leq 10$  range
corresponds to $10.3 \geq B \geq 1.03$ T.

\begin{figure}
\includegraphics{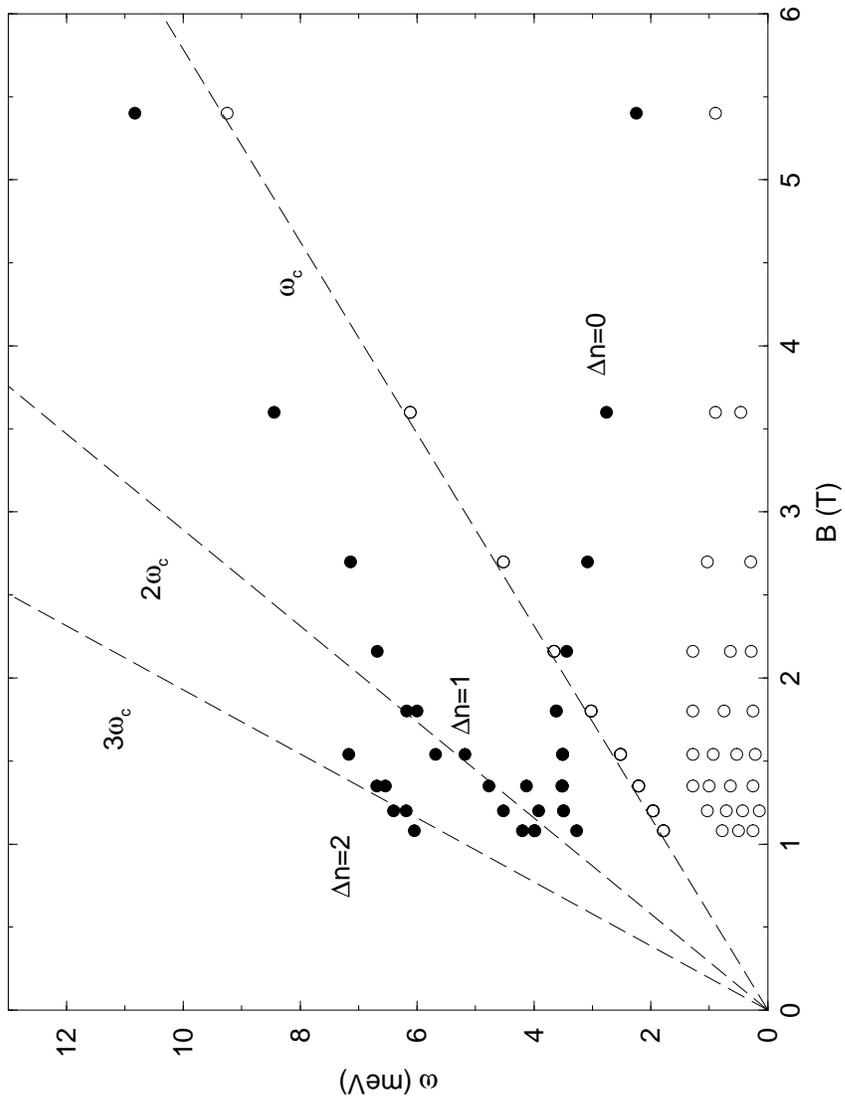}
\vspace*{19.5cm}
\caption[]{
Energies of the more intense antidot CDE's (solid circles) and SDE's
(open circles)
as a function of $B$. From left to right, the calculated points
correspond to $\nu=10$ to 2.
}
\label{fig6}
\end{figure}

\begin{figure}
\includegraphics{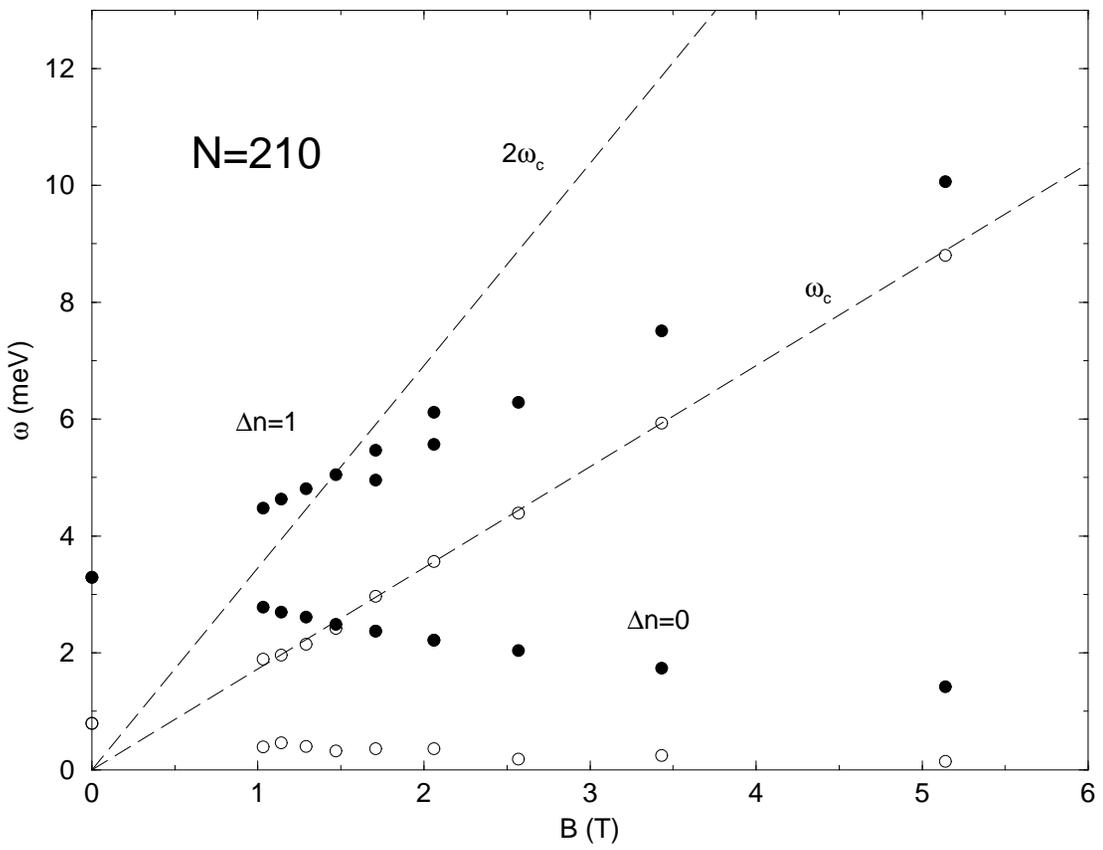}
\vspace*{19.5cm}
\caption[]{
Same as Fig. \ref{fig6} for the $N= 210$ dot. Also drawn are the
$B=0$ values
}
\label{fig7}
\end{figure}

The energies of the more intense CDE's of the 
antidot and of the $N=210$ QD have been drawn as a function of $B$
in Figs. \ref{fig6} and \ref{fig7}, respectively
(a more complete version of Fig. \ref{fig7} can be found in Ref.
\onlinecite{Ser99}). Also drawn are the cyclotron frequency
$\omega_c$, and $2\,\omega_c$ and $3\,\omega_c$ as well. These 
figures  show that the interaction of the bulk
magnetoplasmon with the harmonics of the cyclotron resonance
$n\omega_c$, resembling the Berstein
modes\cite{Gud95}, which causes the splitting of the magnetoplasmon
near $2\,\omega_c$ and $3\,\omega_c$, is well reproduced by the
calculations. Alongside the groups of peaks corresponding 
to CDE's we have indicated the change in radial
quantum number $n$ of the sp wave functions involved in the transition.
This change $\Delta n$ is unambiguosly identified in TDLSDFT 
calculations introducing energy cutoffs in the electron-hole 
energy denominators $\Delta \epsilon$
entering the definition of the free correlation function\cite{Ser99}
$\chi^{(0)}_{\sigma\sigma'}$: limiting the electron-hole pairs to those
having  $\Delta \epsilon < \omega_c$, only  $\Delta n= 0$ peaks
apear in the response, whereas limiting them to those with
$\Delta \epsilon < 2 \,\omega_c$, no $\Delta n= 2$ peaks appear.
These cyclotron-like $\Delta n= 2$ peaks have been experimentally
detected\cite{Hoc99}.

\begin{figure}
\includegraphics{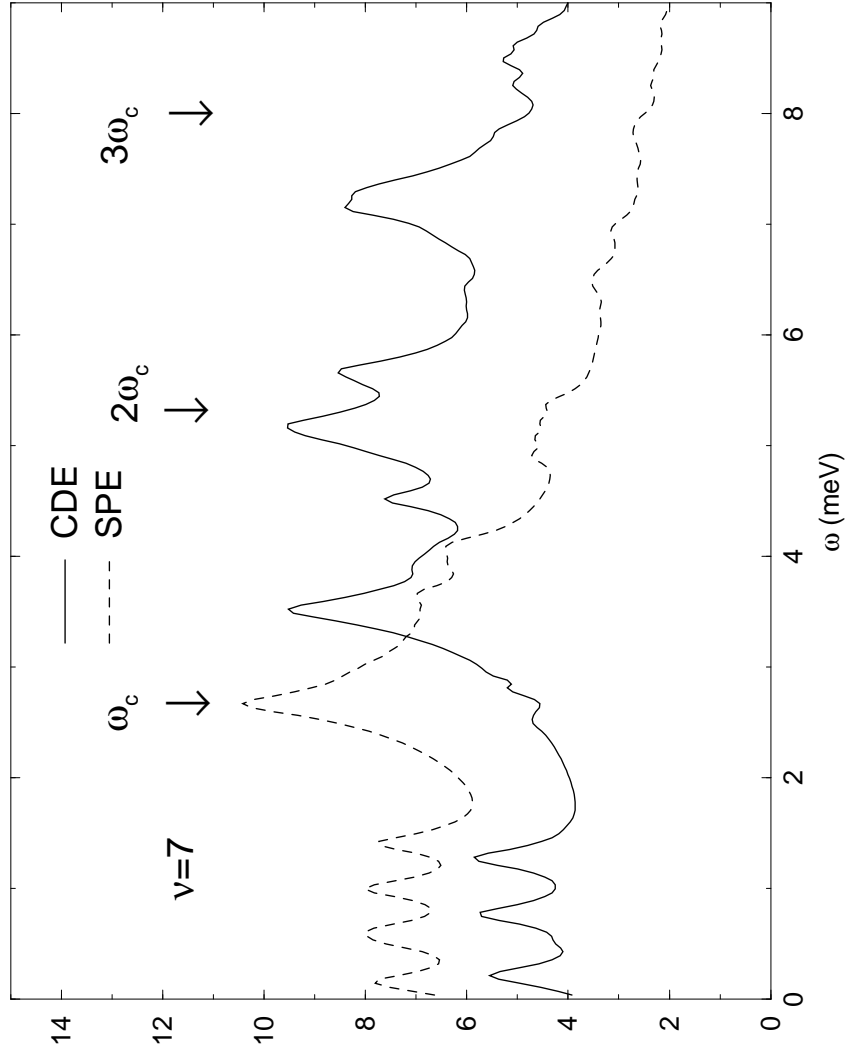}
\vspace*{19.5cm}
\caption[]{
Antidot
CDE (solid line) and SPE (dashed line) strength functions at $\nu=7$
in a logarithmic scale. 
}
\label{fig8}
\end{figure}

The correlations introduced by TDLSDFT have a dramatic effect on the
FIR response of the system which might not always be properly 
recognized. Not only they appreciably shift the CDE energies
with respect to the SPE's but also affect their intensity,
causing that peaks having a sizeable strength in the SPE response
are nearly washed out, whereas others hardly perceived in the SPE
response adquire a large strength. To illustrate it we show in Fig.
\ref{fig8} the single particle and charge density strength 
functions for $\nu=7$ in a logarithmic scale. 
It can be noticed that in the SPE strength the $2\,\omega_c$ peak, which
has an intensity six orders of magnitude smaller than the $\omega_c$
peak  and cannot be seen at the scale
of Fig. \ref{fig3}, emerges as a very collective peak in the CDE
strength, of intensity similar to that of the  bulk
$\Delta n=1$ peak. Similarly, the weak edge SPE's loose
their dimy intensity, transferring it to the very collective
$\Delta n=0$ edge CDE. Of course, in actual TDLSDFT calculations 
all these modes are reciprocally influenced, but the gross
effect is as described.

\begin{figure}
\includegraphics{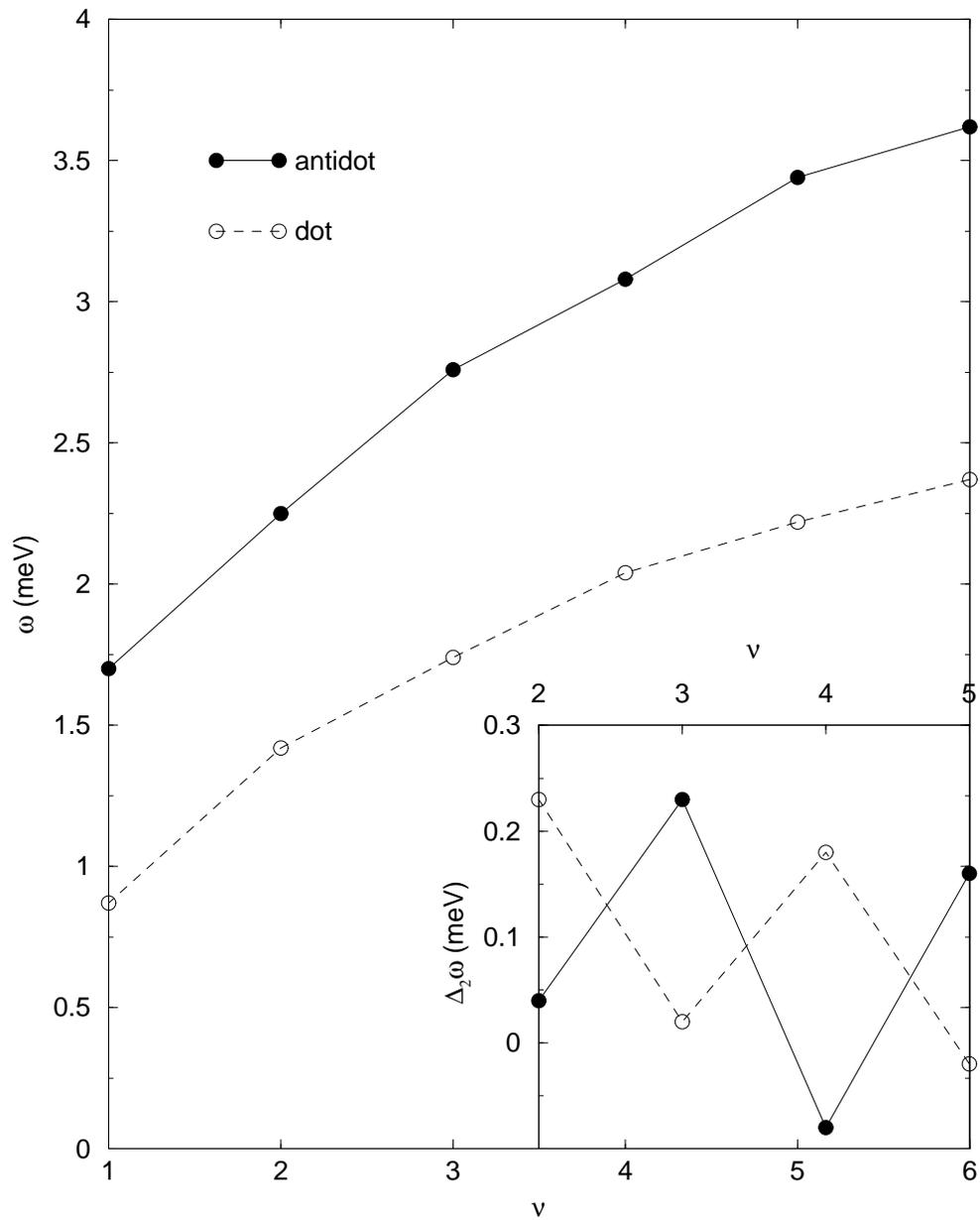}
\vspace*{19.5cm}
\caption[]{
Edge magnetoplasmon energies of the antidot (solid circles) and of
the $N=210$ dot (open circles) as a function of filling factor.
The insert displays the second differences $\Delta_2\, \omega$.
The lines have been drawn to guide the eye.
}
\label{fig9}
\end{figure}

TDLSDFT does reproduce the oscillation of the edge magnetoplasmon energy
as a function of filling factor\cite{Bol96}. 
Figure \ref{fig9} shows the edge magnetoplasmon energy as a
function of $\nu$ for the antidot and the $N=210$ dot. The oscillations
and the out-of-phase effect\cite{Bol96} are tiny effects better
seen plotting the second differences of the edge magnetoplasmon
energy:
\begin{equation}
\Delta_2\, \omega(\nu) = \omega(\nu+1)- 2\,\omega(\nu) + \omega(\nu-1)
\,\,\, .
\label{eq8}
\end{equation}
The energy oscillations are  clearly seen in the
Fig. \ref{fig9} insert, as well as the experimental finding that
maxima in the edge magnetoplasmon energies correspond to half-filled
Landau levels in antidots, and to filled Landau levels in dots.

>From Figs. \ref{fig3} and \ref{fig6} we conclude that for an isolate
antidot, TDLSDFT is unable to produce the  anticrossing
of the edge  and the cyclotron modes at small $B$ values which is
observed in short-period antidot arrays. Yet,
the behavior of the antidot and dot
$\Delta n=0$ edge branch near the crossing is quite distinct,
manifesting a different curvature. The low-$B$ behavior
of the edge brach experimentally found in short-period antidot
arrays is likely influenced by neighboring antidots. Notice 
that ${\cal L} \sim 1.9\, a^*_0$ at $\nu=6$, which is roughly 
half the effective surface width of the antidot lattice period
of Ref. \onlinecite{Zha92}.

Figures \ref{fig3} and \ref{fig6} also show that the TDLSDFT 
antidot response is very fragmented at low $B$. This renders  
tricky to distinguish bulk from edge charge modes. Notice  
that at $\nu=10$ the high energy peak
in Fig. \ref{fig3} is a $\Delta n=2$ mode, whereas the low energy peak
is a mixture of $\Delta n=0$ and $\Delta n=1$ peaks. The situation
worsens increasing $\nu$, which is beyond our capabilities.
For instance, to have results at $B \sim 0.5$ T
one would need to carry out calculations for $M \sim 20$.

\begin{figure}
\includegraphics{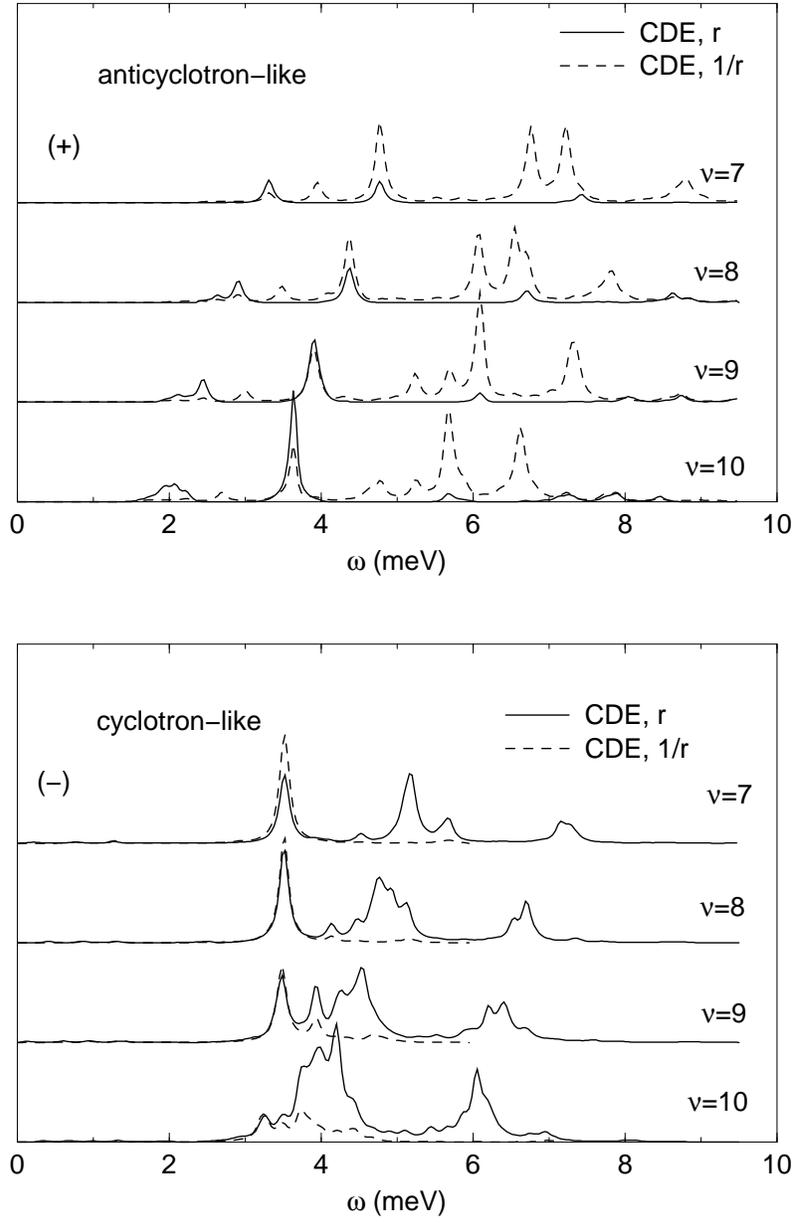}
\vspace*{19.5cm}
\caption[]{
Antidot charge density  strength functions in arbitrary units for 
$\nu=10$ to 7 caused by $r$ (solid lines) and $1/r$ (dashed lines) 
fields. Top panel, anticyclotron-like polarized ($+$) modes.
Bottom panel, cyclotron-like polarized ($-$) modes.
}
\label{fig10}
\end{figure}

A possible way to partially disentangle these modes is to study
the charge density response to the $f(r)= 1/r$ operator. 
Its $r$ dependence makes this very ineffective in
exciting cyclotron-like modes while it is well suited to describe
edge oscillations. This can be seeing in Fig. \ref{fig10}, where
we have plotted the CDE's caused by $r$ and $1/r$ for $\nu=10$ to 7.
In the bottom panel, the intensity of the $1/r$ peaks  has been
multiply by a factor of 50,000 to allow for a sensible
comparison with the $r$ peaks. It can then be seen in that panel 
that both operators excite the same $\Delta L_z = +1$ peaks, 
but with much
different intensity, and that indeed, the low energy peak clearly
visible in the response is an edge peak.

Finally, we have plotted in the top panel of that figure the CDE 
caused by the $e^{-\imath\theta}$ component of the $F$ operators 
in Eq. \ref{eq5}. These are anticyclotron-like excitations of
very weak intensity: to have them at the same scale as
in the bottom panel, one should divide the scale of the $r$ modes
by a factor of 40, and the scale of the $1/r$ modes by a factor of 4,
i.e., the $1/r$ operator excites the cyclotron-like and 
anticyclotron-like main peaks with roughly equal intensity.
In agreement with the experiment (see especially Fig. 1 of Ref.
\onlinecite{Hoc99}),
it can be seen on the one hand that their intensity quickly
decrease as $B$ increases, and on the other hand 
that their energies are very
similar to these of the cyclotron-like modes. This makes these modes
experimentaly accessible only using  circular polarized radiation.

\section{acknowledgments}

It is a pleasure to thank Vidar Gudmundsson
for useful discussions.
This work has been performed under grants PB98-1247 and PB98-0124
from DGESIC, and SGR98-00011 from Generalitat of Catalunya. A.E.
acknowledges support from the Direcci\'on General de Ense\~nanza
Superior (Spain).


\end{document}